\documentclass[journal,twocolumn]{IEEEtran}
\usepackage{amsfonts}
\usepackage{times}
\usepackage{graphicx}
\usepackage{latexsym}
\usepackage{dsfont}
\usepackage{amssymb}
\usepackage{amsmath}
\usepackage{cite}
\usepackage{verbatim}
\usepackage{subfigure}

\newcommand{\figref}[1]{{Fig.}~\ref{#1}}

\def\bb0{{\mathbb{0}}}

\def\ba{{\mathbf{a}}}
\def\bb{{\mathbf{b}}}

\def\bff{{\mathbf{f}}}

\def\bh{{\mathbf{h}}}

\def\bp{{\mathbf{p}}}

\def\b0{{\mathbf{0}}}

\def\bA{{\mathbf{A}}}

\def\bI{{\mathbf{I}}}

\def\cA{\mathcal{A}}

\def\sf0{{\mathsf{0}}}

\usepackage{epstopdf}
\usepackage{enumerate}
\usepackage{algorithmicx}
\usepackage{algorithm}
\usepackage{amsmath}
\usepackage[noend]{algpseudocode}
\usepackage{float}
\usepackage{hyperref}
\usepackage{color}
\usepackage{makeidx}
\usepackage{bbm}
\usepackage{graphicx}
\usepackage{lipsum}

\newcommand{\sref}[1]{{Section}~\ref{#1}}

\newcommand{\pinv}[1]{\ensuremath{#1^{\dagger}}} 	

\begin{document}
\title{DeepMIMO: A Generic Deep Learning Dataset for Millimeter Wave and Massive MIMO Applications}
\author{Ahmed Alkhateeb\\ Arizona State University, Email: alkhateeb@asu.edu}
\maketitle

\begin{abstract}

Machine learning tools are finding interesting applications in millimeter wave (mmWave) and massive MIMO systems. This is mainly thanks to their powerful capabilities in learning unknown models and tackling hard optimization problems. To advance the machine learning research in mmWave/massive MIMO, however, there is a need for a common dataset. This dataset can be used to evaluate the developed algorithms, reproduce the results, set \textit{benchmarks}, and compare the different solutions. In this work, we introduce the DeepMIMO\footnote{The latest versions of the DeepMIMO dataset paper and codes can be found on the dataset website \cite{DeepMIMODataset}.} dataset, which is a generic dataset for mmWave/massive MIMO channels. The DeepMIMO dataset generation framework has two important features. First, the DeepMIMO channels are constructed based on accurate ray-tracing data obtained from Remcom Wireless InSite \cite{Remcom}. The DeepMIMO channels, therefore, capture the dependence on the environment geometry/materials and transmitter/receiver locations, which is essential for several machine learning applications. Second, the DeepMIMO dataset is generic/parameterized as the researcher can adjust a set of system and channel parameters to tailor the generated DeepMIMO dataset for the target machine learning application. The DeepMIMO dataset can then be completely defined by the (i) the adopted ray-tracing scenario and (ii) the set of parameters, which enables the accurate definition and reproduction of the dataset. In this paper, an example DeepMIMO dataset is described based on an outdoor ray-tracing scenario of 18 base stations and more than one million users. The paper also shows how this dataset can be used in an example deep learning application of mmWave beam prediction.
\end{abstract}

\section{Introduction} \label{sec:Intro}

 Millimeter wave (mmWave) and massive MIMO are key enabling technologies for current and future wireless systems \cite{Rappaport2013a,HeathJr2016,Larsson2014a,Alkhateeb2014d,Boccardi2014,Jungnickel2014,Bjoernson2016a,Mumtaz2016}. This is mainly thanks to the very high data rates and multiplexing gains promised by these technologies. Employing large numbers of antennas, however, imposes critical challenges on mmWave and massive MIMO systems in supporting highly mobile users, ensuring reliability, and enabling  low-complexity coordination among others. One main reason for these challenges is the large training, feedback, and coordination overhead associated with the large channel matrices. These channels, however, are intuitively some functions of the environment geometry, building materials, transmitter/receiver locations, etc. This motivated using machine/deep learning tools that leverage low-overhead features of the environment and user setups and learn how to use them to predict mmWave and massive MIMO channels/beams \cite{Alkhateeb2018,Li2018,Wang2018a,Va2017a}, enhance system reliability and proactive hand-off \cite{Alkhateeb2018a,Mismar2018}, and enable low-complexity base station coordination \cite{Alkhateeb2018}.

 \textbf{The Need for a Dataset:}  
 To advance the machine learning research in mmWave and massive MIMO, it is crucial to have sufficiently large dataset that researchers can use for (i) evaluating the performance of their machine learning algorithms, (ii) reproducing the results of the other papers, and (iii) setting \textit{benchmarks} and comparing the different algorithms based on common data. Further, we define the following two important requirements for a useful dataset in mmWave and massive MIMO applications. 
 \begin{itemize}
 	\item \textbf{The dataset channels represent the environment:} 
 	Most of the machine learning applications in mmWave and massive MIMO rely on leveraging the correlation between some features of the environment setup (geometry, materials, transmitter/receiver locations, etc.) and the channels or beamforming vectors  \cite{Alkhateeb2018,Li2018,Alkhateeb2018a,Wang2018a,Va2017a}. Therefore, in order to evaluate these algorithms, the dataset channels should capture this dependence on the environment. To achieve that, the channels need to be either collected via real-world measurements or constructed from accurate ray-tracing data. 
 	
 	\item \textbf{The dataset is generic (parametrized):} 
 	Different than machine learning research in other fields, such as computer vision and natural language processing, that is mainly focused on developing and analyzing machine learning models, a major part of the machine learning research in mmWave/massive MIMO is the pre- and post-processing of the data. Further, it is normally important in wireless communication research to study the performance of the developed solutions under various system/channel scenarios. Therefore, a fixed dataset with a specific system/channel assumptions and a pre-defined set of features will highly limit the machine learning research space in mmWave/massive MIMO. This motivates the development of a \textit{generic} dataset where the researcher can adjust the key system/channel parameters and can do pre- and post-processing on the generated data.  
 \end{itemize} 
 
 While some methodologies for MIMO data generation have been proposed before for mobile applications \cite{Klautau2016,Wen2018}, there is no available dataset for mmWave/massive MIMO that satisfies the mentioned requirements, to the best of our knowledge.

\textbf{The DeepMIMO Dataset:}  In this work, we introduce the channels' dataset, DeepMIMO, which is designed for machine/deep learning research in mmWave and massive MIMO applications. More specifically, using this channels' dataset, researchers can easily construct the inputs and outputs of several machine learning applications. The DeepMIMO dataset generation framework has the following important features.
  \begin{itemize}
 	\item The DeepMIMO channels are constructed from accurate ray-tracing data. These data are obtained from the ray-tracing simulator, Wireless InSite, developed by Remcom \cite{Remcom}. Remcom Wireless InSite \cite{Remcom}, is widely used in mmWave and massive MIMO research at both industry and academia \cite{Va2017a,Alkhateeb2018,Alkhateeb2018a,Khawaja2018}, and has been verified with real-world channel measurements \cite{Li2015a,Wu2016,Khawaja2018}. The DeepMIMO channels constructed using this ray-tracing simulation capture the dependence on the environment geometry/materials as well as the transmitter/receiver locations, which is essential for several machine learning applications in mmWave and massive MIMO systems.  	
 	
 	\item The DeepMIMO dataset is generic (parametrized). More clearly, the DeepMIMO dataset generation framework is designed to generate channel datasets based on a set of parameters that can be adjusted by the researcher. This allows tailoring the DeepMIMO dataset to fit the specific machine learning application of interest. This set of parameters controls various system and channel aspects such as the number of antennas, the number of OFDM subcarriers, and the number of channel paths. 
 	
 	\item The DeepMIMO dataset is simple to define and to generate. More specifically, the DeepMIMO dataset is completely defined by (i) the ray-tracing scenario and (ii) the parameters set. This means that any researcher can easily define the adopted dataset and perfectly generate the same dataset defined in other papers by using the same ray-tracing scenario and parameters set. Further, using the DeepMIMO dataset generation framework to generate the desired dataset is very simple as will be shown in \sref{sec:use}.
 \end{itemize}

The rest of this paper is organized as follows. In \sref{sec:general}, we present the general framework of the DeepMIMO dataset generation process, highlighting the key elements in this framework. Then, a detailed description of these elements is provided in \sref{sec:Explanation}, along with an example Wireless InSite ray-tracing scenario. This example scenario has 18 base stations and more than one million users, which generates a sufficiently large dataset for several mmWave/massive MIMO machine learning applications. In \sref{sec:use}, we describe in detail how to use the DeepMIMO dataset generation code. Finally, in \sref{sec:Application}, we present an example on using the DeepMIMO dataset to construct the inputs/outputs of the machine learning model and generate the mmWave beam prediction results in \cite{Alkhateeb2018}.

\textbf{Notation}: We use the following notation throughout this paper: $\bA$ is a matrix, $\ba$ is a vector, $a$ is a scalar, and $\cA$ is a set. $|\bA|$ is the determinant of $\bA$, $\|\bA \|_F$ is its Frobenius norm, whereas $\bA^T$, $\bA^H$, $\bA^*$, $\bA^{-1}$, $\pinv{\bA}$ are its transpose, Hermitian (conjugate transpose), conjugate, inverse, and pseudo-inverse respectively. $[\bA]_{r,:}$ and $[\bA]_{:,c}$ are the $r$th row and $c$th column of the matrix $\bA$, respectively. $\mathrm{diag}(\ba)$ is a diagonal matrix with the entries of $\ba$ on its diagonal. $\bI$ is the identity matrix and $\mathbf{1}_{N}$ is the $N$-dimensional all-ones vector.
\section{DeepMIMO Dataset: The General Framework} \label{sec:general}

\begin{figure*}[t]
	\centerline{
		\includegraphics[width=1.6\columnwidth]{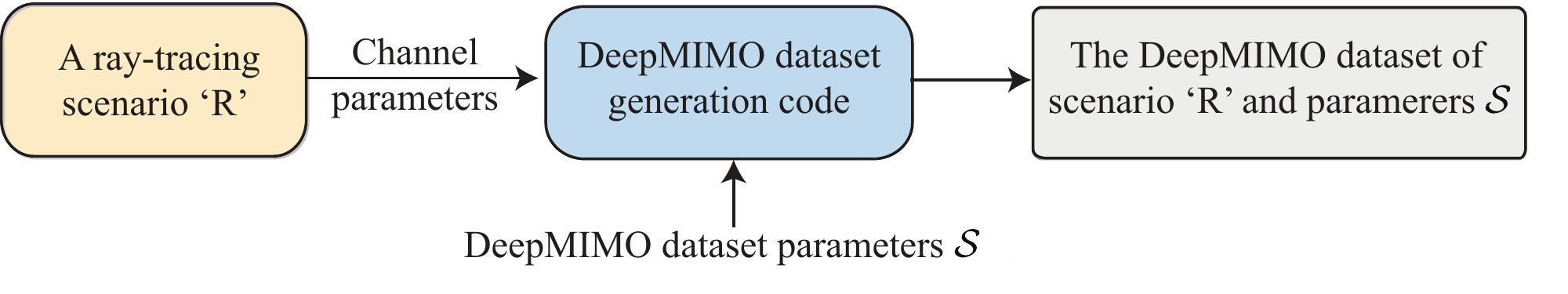}
	}
	\caption{The General framework for generating the DeepMIMO dataset. The DeepMIMO dataset is defined based on (i) the given ray tracing scenario and (ii) the set of dataset parameters $\mathcal{S}$.}
	\label{fig:framework}
\end{figure*}

In this section, we briefly explain the main motivation for having a \textit{generic} MIMO dataset for deep learning applications, and highlight the general framework of our DeepMIMO dataset.
In mmWave and massive MIMO research, the main signal processing tasks, e.g., precoding, channel estimation, beam tracking, and user selection, evolve around the characteristics of the wireless channels.
Some of these characteristics, such as the correlation between the user channels at different locations of the environment, depend heavily on the environment geometry and materials. This makes it hard to generate channels that capture these environment-dependent characteristics using statistical channel models. 
In fact, most of the proposed mmWave and massive MIMO applications for machine learning relied on these environment-dependent channel characteristics to do their functions. Examples include predicting the beamforming and channel matrices based on the user RF signature \cite{Alkhateeb2018,Li2018}, or based on the user location \cite{Wang2018,Va2017a,Asadi2018} and predicting the future blockage based on the sequence of previously selected beams \cite{Alkhateeb2018a}.   
This explains why we need ray-tracing based channel generation in mmWave and massive MIMO based machine learning applications.

The channels generated using ray-tracing simulations capture the geometry-based characteristics, such as the correlation between the channels at different locations, and the dependence on the materials of the various elements of the environment, among others.    
With this motivation, we build a generic (parametrized) dataset for mmWave and massive MIMO channels with the goal of facilitating the deep learning research in this area and enabling results replication and algorithms comparisons. 
Using our generic/parametrized dataset, researchers can tune several parameters, such as the number of antennas, the array configuration, and the number of subcarriers, to craft the dataset that fits their application. 
The general framework for our DeepMIMO dataset generation is illustrated in \figref{fig:framework}. Next, we highlight the main elements in this framework. 
\begin{itemize}
	\item \textbf{The ray-tracing scenario `R':} 
	The ray-tracing scenario consists of a number of base stations (or access points) and users geographically distributed in a certain outdoor or indoor environment. Typically, in the ray-tracing scenario, the base stations and users have omni or quasi-omni antennas. The outputs of the ray-tracing simulation include the channel parameters (angles or arrival/departure, path gains, etc.) for the channels between every transmitter and receiver. In our dataset, we use the accurate ray-tracing simulator, Wireless InSite by Remcom \cite{Remcom} to obtain the ray-tracing outputs. These outputs for some ray-tracing scenarios are available on our dataset website \cite{DeepMIMODataset}. Further, to have a large enough dataset for deep learning applications, our ray-tracing scenarios  include a large number of base stations and users. An example on these scenarios is explained in detail in \sref{subsection:scenario}.
	
	\item \textbf{The dataset parameters $\mathcal{S}$:}
	Since different machine learning applications require different datasets, we designed DeepMIMO as a parametrized dataset. This allows the researchers to adjust a set of parameters, $\mathcal{S}$, in the dataset generation code to generate a  dataset that is customized for their application. This achieves two main objectives: (i) it provides the researchers with a wide control over the system setup and the antenna configuration, and (ii) it facilitates results reproducibility, as the researchers just need to state the parameters set, $\mathcal{S}$, and the adopted ray-tracing scenario, `R', to completely define the generated dataset . We describe these parameters in detail in \sref{subsection:parameters}. 
	
	\item \textbf{The DeepMIMO dataset generation code:} 
	Given the channel parameters generated from the ray-tracing scenario `R', and based on the selected parameters set $\mathcal{S}$,  the DeepMIMO dataset generation code will construct the channel matrices for all the selected transmitter-receiver pairs. In addition to the channel matrices, the DeepMIMO dataset includes other important features, such as the user location, which can be leveraged in the machine learning modeling. The DeepMIMO dataset generation code and the structure of the generated dataset are explained in detail in Sections \ref{subsec:datasetGeneration}-\ref{subsec:datasetStructure}.
\end{itemize}

\noindent \textbf{It is important to note that the generated DeepMIMO dataset is completely defined by (i) the adopted ray-tracing scenario `R' and (ii) the dataset parameters set $\mathcal{S}$.} This allows the researchers to easily define their dataset, reproduce the results in other papers, and compare the performance of different algorithms using a common dataset. Next, we explain the DeepMIMO dataset framework in detail in \sref{sec:Explanation}, before showing how we can use the dataset is some mmWave deep learning applications in \sref{sec:Application}.

\section{DeepMIMO Dataset: A Detailed Description} \label{sec:Explanation}

In this section, we will describe in detail the different elements of the DeepMIMO dataset generation process, illustrated in \figref{fig:framework}.  As discussed in \sref{sec:general}, the DeepMIMO dataset is completely defined by the ray-tracing scenario and the set of parameters $\mathcal{S}$. Therefore, to generate the dataset, the researcher will first choose one of the ray-tracing scenarios that are available for the DeepMIMO dataset on the dataset website \cite{DeepMIMODataset}. For each scenario, we provide the channel parameters of every transmitter/receiver pair, which is the first input to the dataset generation code, as shown in \figref{fig:framework}. Then, the researcher will adjust the dataset parameters $\mathcal{S}$ to fit the desired application. Finally, the dataset generation code will construct the DeepMIMO dataset based on the channel parameters  of the adopted ray-tracing scenario and system parameters. In the next few subsections, we describe these aspects in more detail, which is important for leveraging the DeepMIMO dataset and understanding its  full capabilities.

\begin{figure*}[p] 
	\centering
	\includegraphics[width=2\columnwidth]{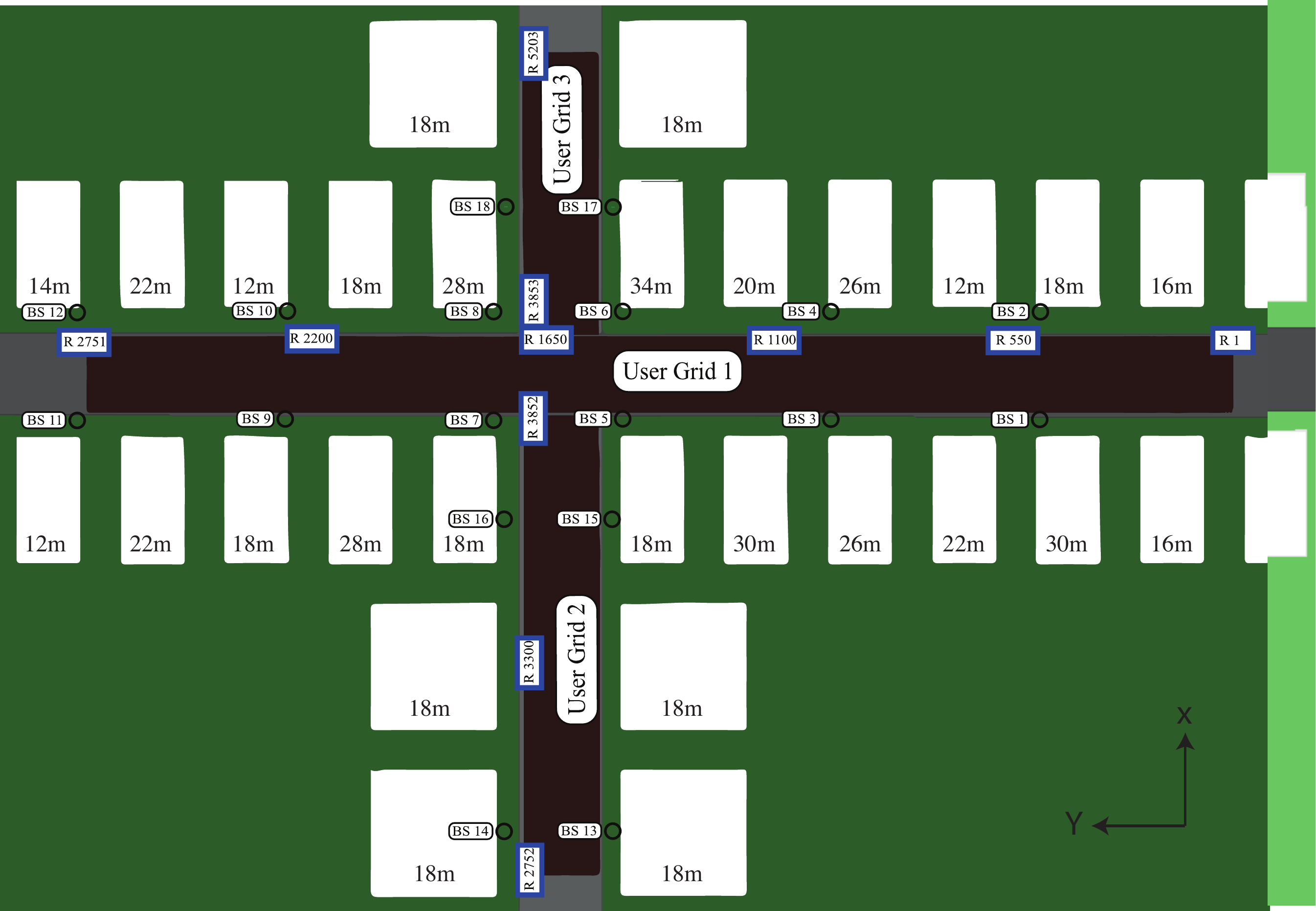}
	\caption{A top view of the 'O1' ray-tracing scenario, showing the two streets, the buildings, the 18 base stations, and the user x-y grids. This ray-tracing scenario is generated using Remcom Wireless InSite \cite{Remcom}. } 	\label{fig:top}
	\bigskip 
	\includegraphics[width=2\columnwidth]{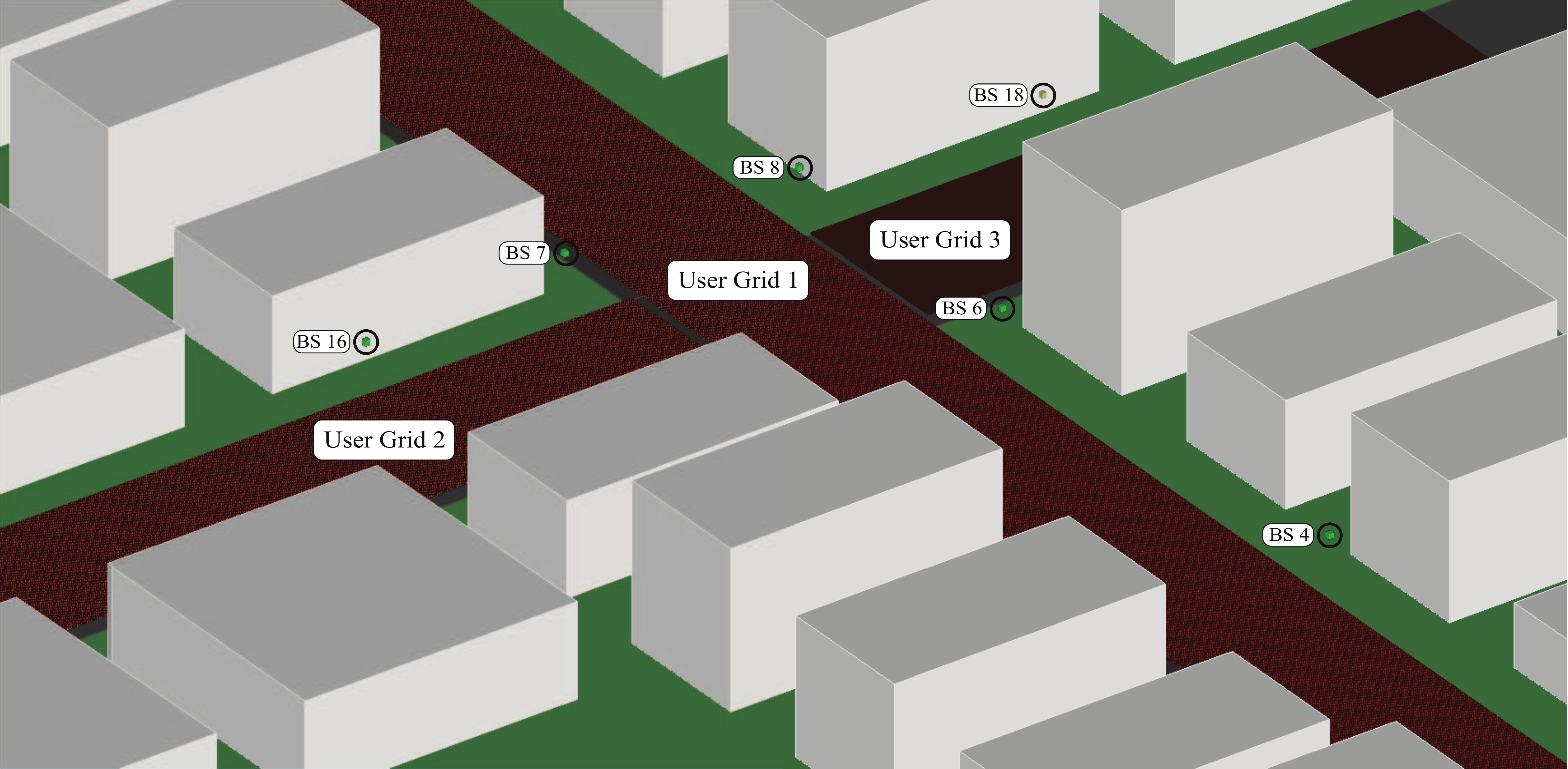}
	\caption{A bird's-eye view of a section of the 'O1' ray-tracing scenario, showing the intersection of the two streets. This ray-tracing scenario is generated using Remcom Wireless InSite \cite{Remcom}. \label{fig:side} }
\end{figure*}

\subsection{Ray-Tracing Scenarios} \label{subsection:scenario}

As described briefly in \sref{sec:general}, the ray-tracing simulations generate channel parameters that capture the dependence on the environment geometry, materials, transmitter/receiver locations, etc., which is crucial for the machine learning applications. In our dataset, the channel parameters are generated using the accurate ray-tracing simulator Wireless InSite by Remcom \cite{Remcom}. These channel parameters are the first inputs to the DeepMIMO dataset generation code as shown in \figref{fig:framework}. On the DeepMIMO dataset website \cite{DeepMIMODataset}, the channel parameters for some ray-tracing scenarios will be available. In this section, we explain in detail one of those scenarios. Understanding these ray-tracing scenarios is important for several reasons: (i) the dataset user needs this understanding to be able to adjust the parameters in the set $\mathcal{S}$, such as the antenna configuration and orientation, given the adopted system and machine learning models, (ii) this understanding of the ray-tracing scenario enables the researcher to develop good explanation of the machine learning outcomes. Now, we explain one ray-tracing scenario, that we call `O1', in detail.

\textbf{The ray-tracing scenario `O1':}  
This is an outdoor scenario of two streets and one intersection, with the top-view shown in \figref{fig:top}. The main street (the horizontal one) is 600m long and 40m wide, and the second street (the vertical one in \figref{fig:top}) is 440m long and 40m wide. In the following bullets, we describe the key components of this ray-tracing scenario.   
\begin{itemize}
	\item \textbf{Base stations:} As shown in \figref{fig:top}, the `O1' ray-tracing scenario includes 18 base stations, BS1-BS18, distributed on both sides of the two streets. The main street has 12 BSs, 6 on each side. The separation between BS1, BS3, and BS5 (or equivalently BS2, BS4, BS6) is constant and equals 100m (and similarly for BS7, BS8, BS9 and BS8, BS10, BS12). For the second street, it includes 6 BSs. The separation between BS13, BS 15, and BS17 (or equivalently BS14, BS16, and BS18) is 150m. The hight of all the BSs is 6m. Further, in the ray-tracing simulation, each BS has only a single half-wave dipole with the axis of the dipole antenna in the $z$-direction. In Sections \ref{subsection:parameters}-\ref{subsec:datasetGeneration}, we will show how the output of this ray-tracing simulation can be used to generate channels for larger antenna arrays at the BSs.
	
	\item \textbf{Users:} Since machine learning applications normally require large training dataset, the `O1' ray-tracing scenario is designed to include more than one million users (exactly, 1,184,923 users). The users are placed in 3 uniform x-y grids as shown in \figref{fig:top}. The first user grid is located along the main street, with a length of 550m  and a width of 35m. It starts from the right on \figref{fig:top}, after 15m of the street starting point, and ends on the left, 35m before the street ending point. This first grid includes 2751 rows, R1 to R2751, with each row having 181 users. The spacing between every two adjacent users in this uniform x-y grid is 20 cm.  The second grid is located in the southern side of the second street, as shown in \figref{fig:top}. It includes 1101 rows, R2752 to R3852, with 181 users at every row. Similar to the first grid, The spacing between every two adjacent users in the second x-y grid is 20 cm. The third grid is located on the northern part of the second street, as shown on \figref{fig:top}. It includes 1351 rows, from R3853 to R5203, with 361 users in every row. Different than the other two grids, the spacing between every two adjacent users in the third x-y grid is 10 cm. Note that the main advantage of having some differences between the three grids is to enable testing different scenarios for the ray-tracing simulations and machine learning applications. Finally, all the users are equipped with a single dipole antenna, with the axis aligning with the z-direction. 

	\item \textbf{Buildings:} In this ray-tracing scenario, the two streets have buildings on both sides. For simplicity, all the building are assumed to be solid, with rectangular shapes. Along the main street, all the buildings have bases of the same dimensions, 30m $\times$ 60m. In the second street, the buildings have bases of 60m $\times$ 60m. The heights of the buildings are different, and the height of every building is written on it in \figref{fig:top}.
	
	\item \textbf{Materials:} In the `O1' ray-tracing scenario, we emulate a $60$ GHz signal propagation setup. Therefore, this scenario adopts the  ITU dry earth 60 GHz material for the two streets and ITU layered drywall 60 GHz for the buildings. These materials are available in the Wireless InSite ray-tracing simulator \cite{Remcom}.     
\end{itemize}

\textbf{Ray-tracing outputs:} 
In the Wireless InSite ray-tracing simulator \cite{Remcom}, we used the X3D model that is developed by Remcom and is capable of providing a highly-accurate 3D propagation model. Further, for simplicity, we considered only the first 4 reflections. For every transmitter-receiver pair, this ray-tracing simulation shoots hundreds of rays in all directions from the transmitter and record the strongest $25$ paths from those that made their ways to the receiver, where strongest paths are the paths with the highest receive power. Further, for every base station-user pair, the specified ray-tracing simulator calculates the channel parameters for every channel. More specifically, for every BS $b$ and user $u$, and for every channel path $\ell$, the ray-tracing simulations outputs (1) the azimuth and elevation angles of departure (AoDs) from the base station, $\phi^{b,u}_\mathrm{az}, \phi^{b,u}_\mathrm{el}$, (2) the azimuth and elevation angles of arrival (AoAs) at the user, $\theta^{b,u}_\mathrm{az}, \theta^{b,u}_\mathrm{el}$, (3) the path receive power, $P_\ell^{b,u}$, (4) the path phase, $\vartheta_\ell^{b,u}$, (5) the propagation delay of the path $\tau_\ell^{b,u}$. In \sref{subsec:datasetGeneration}, we show how these parameters can be used to construct the channel matrix between base station $b$ and user $u$. In addition to the channel parameters, the ray-tracing outputs include the x-y-z location of the user, which can be leveraged as an input feature for machine/deep learning applications. 

\subsection{DeepMIMO Dataset Parameters} \label{subsection:parameters}
In \sref{subsection:scenario}, we described the ray-tracing outputs, which are the first input to the DeepMIMO dataset generation code in \figref{fig:framework}. In order to construct the channel matrices and build the dataset, though, we still need to define the DeepMIMO dataset parameters, such as the number of antennas and antenna spacing, which are the second input to the DeepMIMO dataset generation code. The objective of these dataset parameters, $\mathcal{S}$, is to give the researcher some flexibility in adjusting the DeepMIMO dataset to fit the desired application, which makes it a \textit{generic} dataset. In the following bullets, we list the DeepMIMO dataset parameters.   
\begin{itemize}
	\item \textbf{Active BSs} (defined by \texttt{active$\_$BS} in the MATLAB code): Here, we specify the BSs that we want to activate in the dataset, i.e., the BSs that we want the DeepMIMO dataset generation code to generate the channels connecting them and the mobile users. Specifying the active BSs helps reducing the size of the dataset by focusing on a certain subset of the available BSs. For example, the `O1' ray-tracing scenario includes 18 BSs. If our application requires only the channels between BSs 3, 4, 5, 6 (in \figref{fig:top}) and the mobile users, we set  \texttt{active$\_$BS=[3,4,5,6]}. 
	
	\item \textbf{Active users} (defined by \texttt{active$\_$user$\_$first} and \texttt{active$\_$user$\_$last} in the MATLAB code): Similar to the BSs, we can activate a certain group of users for the DeepMIMO dataset generation code. We do that by specifying the first and last row of the group of users. For example, we can activate the user group from row R1000 to row R1500 by setting \texttt{active$\_$user$\_$first=1000} and  \texttt{active$\_$user$\_$last=1500}.

	\item \textbf{Number of BS antennas} (defined by \texttt{num$\_$ant$\_$x}, \texttt{num$\_$ant$\_$y}, and \texttt{num$\_$ant$\_$z} in the MATLAB code): These parameters specify the number of BS antennas in the x, y, and z axes, assuming a uniform array. Note that the axes are w.r.t. to the ray-tracing axes. For example, considering BS 3 in the `O1' ray-tracing scenario, in \figref{fig:top}. If this BS has a $16 \times 16$ uniform planar array (UPA) along the street, i.e., a UPA in the y-z plane, we set \texttt{num$\_$ant$\_$x=1}, \texttt{num$\_$ant$\_$y=16}, and \texttt{num$\_$ant$\_$z=16}. 
	
	\item \textbf{Antenna spacing} (defined by \texttt{ant$\_$spacing} in the MATLAB code): This parameter specifies the spacing between the elements of the BS antenna array relative to the wavelength. For a half-wavelength antenna spacing, we set  \texttt{ant$\_$spacing=.5}.

	\item \textbf{System bandwidth} (defined by \texttt{bandwidth} in the MATLAB code): This parameter defined the system bandwidth in GHz. For example, for 500 MHz bandwidth, we set \texttt{bandwidth=.5}.
	
	\item \textbf{OFDM parameters} (defined by \texttt{num$\_$OFDM}, \texttt{OFDM$\_$sampling$\_$factor}, and \texttt{OFDM$\_$limit} in the MATLAB code):
	These  parameters specify the number of OFDM subcarriers and at which subcarriers we want the DeepMIMO dataset generation code to calculate the channels. Calculating the channels only at a specific set of subcarriers helps reducing the dataset size. To do that, two parameters \texttt{OFDM$\_$sampling$\_$factor}, and \texttt{OFDM$\_$limit} can be leveraged. While the first parameter, \texttt{OFDM$\_$sampling$\_$factor}, provides the option to consider only a sampled version of the OFDM subcarriers,  the second one,  \texttt{OFDM$\_$limit}, specifies how many \textit{sampled} subcarriers we want to consider. For example, consider an OFDM system with $1024$ subcarriers, if we want to calculate the channels only at the first $64$ subcarriers, we set  \texttt{num$\_$OFDM=1024}, \texttt{OFDM$\_$sampling$\_$factor=1}, and  \texttt{OFDM$\_$limit=64}. Also, considering the same OFDM system, if we want to calculate the channels only at the first 64 sampled subcarriers with a downsampling factor of 4, i.e., at subcarriers 1, 5, 9, ..., 256, then we set \texttt{num$\_$OFDM=1024}, \texttt{OFDM$\_$sampling$\_$factor=4}, and \texttt{OFDM$\_$limit=64}.	 
	
	\item \textbf{Number of channel paths} (defined by \texttt{num$\_$paths} in the MATLAB code): As described in \sref{subsection:scenario}, the ray-tracing simulation outputs the AoA, AoD, etc. of up to $25$ paths for the channel between every BS and user. These paths are ordered according to their received power. For example, the first path is the one with the highest received power. In some applications, we may be interested in considering only the strongest path or the first few paths. To provide this flexibility, we use the parameter \texttt{num$\_$paths} . For example, if we want to consider only the strongest 3 paths, we set \texttt{num$\_$paths=3}. 
\end{itemize}

\subsection{DeepMIMO Dataset Construction Code} \label{subsec:datasetGeneration}

Given the ray-tracing simulation outputs, described in \sref{subsection:scenario}, and the DeepMIMO dataset parameters, explained in \sref{subsection:parameters}, the DeepMIMO dataset generation code constructs the channels between the specified BSs and users. More specifically, consider a DeepMIMO dataset parameters $\mathcal{S}$ that specifies (i) a  number of BS antennas $M=M_xM_yM_z$ with $M_x, M_y, M_z$ the number of BS antennas in the $x, y,$ and $z$ directions, (ii) antenna spacing $d$, (iii) system bandwidth $B$, (iv) a set of OFDM subcarriers $\mathcal{K}$ at which the channels need to be calculated, and (v) a number of channel paths $L$. Then, the DeepMIMO dataset generation code constructs the $M \times 1$ channel vector $\bh_{k}^{b,u}$ for every active BS $b$ and active user $u$, and on each subcarrier $k \in \mathcal{K}$, where $\bh_{k}^{b,u}$ is expressed as
\begin{equation}
\bh_k^{b,u}= \sum_{\ell=1}^{L} \sqrt{\frac{\rho_\ell}{K}} e^{j \left(\vartheta_\ell^{b,u}+ \frac{2 \pi k}{K}  \tau_\ell^{b,u}  B \right)} \ba(\phi^{b,u}_\mathrm{az}, \phi^{b,u}_\mathrm{el}),
\end{equation}
where $\ba(\phi^{b,u}_\mathrm{az}, \phi^{b,u}_\mathrm{el})$ is the array response vector of the BS, and is defined as 
\begin{equation}
\ba\left(\phi^{b,u}_\mathrm{az}, \phi^{b,u}_\mathrm{el}\right)=\ba_z\left(\phi^{b,u}_\mathrm{el}\right) \otimes  \ba_y\left(\phi^{b,u}_\mathrm{az}, \phi^{b,u}_\mathrm{el}\right)  \otimes \ba_x\left(\phi^{b,u}_\mathrm{az}, \phi^{b,u}_\mathrm{el}\right),
\end{equation}
with $\ba_x(.), \ba_y(.), \ba_z(.)$ the BS array response vectors in the $x, y,$ and $z$ directions, and are expressed as 
\begin{align}
\ba_x\left(\phi^{b,u}_\mathrm{az}, \phi^{b,u}_\mathrm{el}\right)&=\left[1, e^{j kd \sin(\phi^{b,u}_\mathrm{el}) \cos(\phi^{b,u}_\mathrm{az})}, ...\right.  \nonumber \\
&\hspace{20pt}\left. ...,e^{j kd (M_x-1)\sin(\phi^{b,u}_\mathrm{el}) \cos(\phi^{b,u}_\mathrm{az})} \right]^T,\\
\ba_y\left(\phi^{b,u}_\mathrm{az}, \phi^{b,u}_\mathrm{el}\right)&=\left[1, e^{j kd \sin(\phi^{b,u}_\mathrm{el}) \sin(\phi^{b,u}_\mathrm{az})}, ...\right.  \nonumber \\
&\hspace{20pt}\left. ...,e^{j kd (M_x-1)\sin(\phi^{b,u}_\mathrm{el}) \sin(\phi^{b,u}_\mathrm{az})} \right]^T, \\
\ba_z\left(\phi^{b,u}_\mathrm{el}\right)&=\left[1, e^{j kd \cos(\phi^{b,u}_\mathrm{el}) }, ...,e^{j kd (M_x-1)\cos(\phi^{b,u}_\mathrm{el}) } \right]^T. 
\end{align}

In addition to the set of OFDM channel vectors $\bh_{k}^{b,u}, k \in \mathcal{K}$ for every active BS $b$ and user $u$, the DeepMIMO dataset also outputs the location of the user $\bp^u=[p_x,p_y,p_z]$ in the x-y-z space. This location information is important as it can be used, for example, as an input feature in some ML applications. 

\textbf{Remark} \textit{ It is important to emphasize here that, as described in this section, the constructed DeepMIMO dataset is a function of only two inputs: (i) the considered ray-tracing scenario `R', and (ii) the dataset parameters set $\mathcal{S}$. This makes it simple for researchers to describe the adopted dataset, reproduce the datasets in the other papers, and compare their proposed designs and algorithms. }

\subsection{Structure of the DeepMIMO Dataset}\label{subsec:datasetStructure}

For every active BS $b$ and user $u$, the DeepMIMO dataset includes (i) the channel vector $\bh_{k}^{b,u}$ for the specified set of subcarriers $\mathcal{K}$, stored as an $M \times |\mathcal{K}|$ matrix, in which the $k$th column represents the channel at the $k$th specified subcarrier, and (ii) the user location $\bp_u$. 

When running the DeepMIMO dataset generation code, it outputs all these data as a one MAT file, named DeepMIMO$\_$dataset.mat.  This file includes only one cell array, called \texttt{DeepMIMO$\_$dataset}. Accessing the data (channels and location) of every base station-user pair is done as follows.
\begin{itemize}
	\item  \texttt{DeepMIMO$\_$dataset$\{\bar{b}\}$.user$\{\bar{u}\}$.channel} accesses the $M \times |\mathcal{K}|$ channel matrix between the active base station $\bar{b}$ and user $\bar{u}$. Note that $\bar{b}$ and $\bar{u}$ represent the $\bar{b}$th active BS and the $\bar{u}$th active user. For example, if we specified the active BSs \texttt{active$\_$BS=[3,4,5,6]} and the active users rows  \texttt{active$\_$user$\_$first=1000} and  \texttt{active$\_$user$\_$last=1500} in the parameters set $\mathcal{S}$, then \texttt{DeepMIMO$\_$dataset$\{1\}$.user$\{1\}$.channel} accesses the channel between the first active BS, which is BS 3, and the first active user, which is the first user in the row R$1000$. 
	\item  \texttt{DeepMIMO$\_$dataset$\{\bar{b}\}$.user$\{\bar{u}\}$.loc} accesses the position vector $\bp_{\bar{u}}$ of the $\bar{u}$th active user. 
\end{itemize} 

In the following section, we describe in detail how to use the DeepMIMO dataset. 

\section{How to Use the DeepMIMO Dataset?} \label{sec:use}

Given the framework of the DeepMIMO dataset generation in \figref{fig:top}, the researcher will need to define the ray-tracing scenario and the parameters set in order to generate the DeepMIMO dataset that is tailored for the desired application. More specifically, the steps for generating the DeepMIMO dataset can be summarized as follows. 
\begin{enumerate}
	\item From the DeepMIMO dataset website \cite{DeepMIMODataset}, download the 'DeepMIMO generation code' file and expand/ uncompress it.
	\item From the DeepMIMO dataset website \cite{DeepMIMODataset}, download the ray-tracing output files for the adopted scenario, for example the 'O1' scenario, and expand it. Note that the 'O1' ray-tracing scenario is described in detail in \sref{subsection:scenario}.
	\item Add the folder of the ray-tracing scenario, for example the 'O1' folder, to the path 'DeepMIMO Dataset Generation/RayTracing Scenarios/'.
	\item Open the  file 'DeepMIMO$\_$Dataset$\_$Generation.m' and adjust the DeepMIMO dataset parameters.  Note that these parameters are described in detail in \sref{subsection:parameters}. 
	\item From the MATLAB command window, call the function  [DeepMIMO$\_$dataset]=DeepMIMO$\_$Datase$\_$Generator(). This function will generate the DeepMIMO dataset given the defined ray-tracing scenario and adopted parameters set. 
	\item Given the generated DeepMIMO dataset, the channels and users locations can be accessed as described in \sref{subsec:datasetStructure}.
\end{enumerate}

In the following section, we show an example of how to use the DeepMIMO dataset in one machine learning application.

\section{An Example Application: Beam Prediction} \label{sec:Application}
The DeepMIMO dataset includes a large number of channel matrices that have geometric meaning, as they are generated based on a ray-tracing model. In this section, we will show how this dataset can be utilized in a deep learning mmWave application. More specifically, we will use the DeepMIMO dataset to evaluate the performance of the deep learning coordinated beamforming algorithm proposed in \cite{Alkhateeb2018}. In this beamforming algorithm, a machine learning model uses the uplink pilot received with omni-antennas at multiple BSs to predict the best beamforming vector at each one of the coordinating BSs. Next, we will define the adopted DeepMIMO dataset before showing how it can be used in our mmWave beam prediction application. [all codes are available in \cite{DeepMIMODataset}.]

\begin{table}[h]
	\caption{The adopted DeepMIMO dataset parameters}
	\begin{center}
		\begin{tabular}{ | c | c | }
			\hline
			\textbf{DeepMIMO Dataset Parameter} & \textbf{Value} \\ \hline \hline
			Active BSs 				&    3, 4, 5, 6  \\ \hline
			Active users 		   &  From row R1000 to row R1300  \\ \hline
			Number of BS Antennas & $M_x=1, M_y=32, M_z=8$ \\ \hline
			Antenna spacing & d=0.5 \\ \hline 
			System bandwidth & B=0.5 GHz \\ \hline 
			Number of OFDM subcarriers & 1024 \\ \hline
			OFDM sampling factor & 1 \\ \hline
			OFDM limit & 64 \\ \hline 
			Number of paths & 5 \\ \hline
		\end{tabular}
	\end{center}
	\label{tabel:DeepMIMO_parameters}
\end{table}

\textbf{Dataset definition:} One main advantage of the DeepMIMO dataset is that it is completely defined by the ray-tracing scenario and the parameters set. This makes it easy for researchers to define their dataset, reproduce the dataset in other papers, and compare/benchmark their algorithms. In our simulations, we consider the DeepMIMO dataset with the `O1' ray-tracing scenarios and with the parameters set in Table \ref{tabel:DeepMIMO_parameters}.

\textbf{Constructing the deep learning  dataset for \cite{Alkhateeb2018}:}
The deep learning coordinated beamforming algorithm in \cite{Alkhateeb2018} adopts a supervised learning model to learn the mapping between the OFDM omni-received sequence at a number of BSs (in our example, 4 BSs) and the beamforming vector at every one of them. Every data point in the dataset that trains this deep learning model consists then of (i) the input which is the omni-received OFDM sequence at 4 BSs, and (ii) the output which is the achievable rate of the candidate beamforming vectors. 

\begin{figure}[t]
	\centerline{
		\includegraphics[width=1.1\columnwidth]{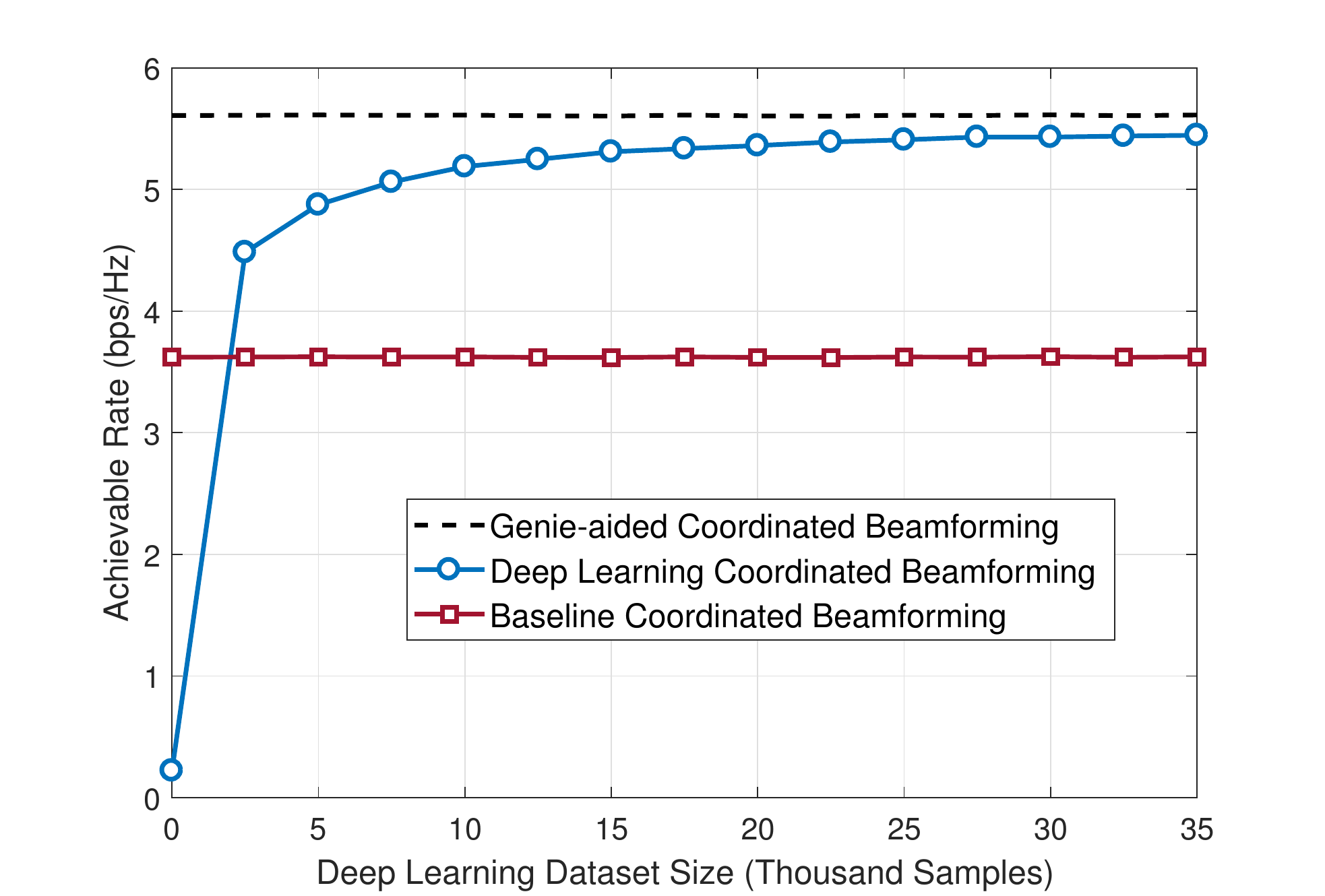}
	}
	\caption{The achievable rate of the deep learning coordinated beamforming algorithm, \cite{Alkhateeb2018}, versus different sizes of training sets. This figure is generated using the DeepMIMO dataset.}
	\label{fig:ML_result}
\end{figure}

\begin{figure}[h]
	\centerline{
		\includegraphics[width=.8\columnwidth]{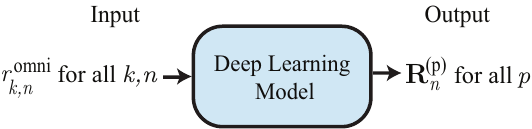}
	}
	\caption{The supervised deep learning model in \cite{Alkhateeb2018} that learns the mapping from the omni-received sequence collected from a number of BSs, $r_{k,n}^\mathrm{omni} \forall k,n$, and the achievable rate with every candidate beamforming vector $R^{(p)}_n, \forall p$ at one of the coordinating BSs $n$.}
	\label{fig:ML_Model}
\end{figure}

Given the DeepMIMO dataset, we can generate these inputs and outputs for every user $u$ as follows. 
\begin{itemize}
	\item  For every BS $n$ (of the active BSs 3, 4, 5, 6), and for every subcarrier $k$ (of the considered set of subcarriers), we can access the channel vector $\bh^{n,u}_{k}$ from the DeepMIMO dataset as described in \sref{subsec:datasetStructure}. Then, we obtain the input to the deep learning model as $r^\mathrm{omni}_{k,n}=\left[\bh^{n,u}_{k}\right]_{1}$, i.e., the first element of the vector. 
	\item For every BS $n$ (of the active BSs 3, 4, 5, 6), the achievable rate, $R_n^{(p)}$, when the candidate beamforming vector $\bff_p$ is adopted is calculated as
	\begin{equation}
			R_n^{(p)}=\frac{1}{\left|\mathcal{K}\right|} \sum_{k \in |\mathcal{K}|} \log_2 \left(1+\mathsf{SNR}   \left|  \bff_p^T  \bh_{k}^{n,u}     \right|^2     \right), 
	\end{equation} 
	where the channel vector $ \bh_{k}^{n,u}$ is obtained from the DeepMIMO dataset. We refer the readers to \cite{Alkhateeb2018} for more details on the beamforming algorithm and the deep learning model. Further, the MATLAB and python codes that implement the beamforming algorithm and machine learning model based on the DeepMIMO dataset is available at \cite{DeepMIMODataset}.
\end{itemize}

\textbf{Simulation results:} With the generated deep learning data points that is based on the DeepMIMO dataset, we trained the deep learning model that is described in \cite{Alkhateeb2018} to obtain the performance results in \figref{fig:ML_result}. We encourage the researchers to reproduce these results and use them for comparisons with their proposed algorithms. The codes to generate \figref{fig:ML_result} are available on the DeepMIMO dataset website \cite{DeepMIMODataset}.

\section{Acknowledgment}
The author thanks Mr. Tarun Chawla and Remcom \cite{Remcom} for supporting and encouraging this work. The author also thanks Xiaofeng Li, Muhammad Alrabeiah, and Abdelrahman Taha for their valuable feedback and suggestions.

\section{Conclusion}
In this paper, we presented the DeepMIMO dataset which is a channels' dataset designed to advance machine learning research in mmWave and massive MIMO systems. The DeepMIMO dataset generation framework constructs the MIMO channels based on ray-tracing data obtained from the accurate ray-tracing simulation, Remcom Wireless InSite \cite{Remcom}. The DeepMIMO channels, therefore, capture the dependence on the various elements of the environment such as the scatterers geometry and transmitter/receiver locations, which is important for machine learning research. Further, the DeepMIMO dataset was designed to be generic, which enables the researcher to generate the dataset based on adjustable system/channel parameters. We also described an example DeepMIMO dataset based on a ray-tracing scenario of 18 base stations and more than one million users. Further, we explained how to use the DeepMIMO dataset generation code to adjust the set of parameters and generate the dataset. Finally, as an example application, we showed how the DeepMIMO dataset can be used to construct the inputs/outputs of the deep learning model of the mmWave deep learning coordinated beamforming solution \cite{Alkhateeb2018}.


\end{document}